\documentclass[twocolumn]{aastex63}
\hypersetup{linkcolor=red,citecolor=blue,filecolor=cyan,urlcolor=black}

\newcommand{\kms}{km\,s$^{-1}$}
\newcommand{\OIII}{O\,{\scriptsize III}}
\newcommand{\OII}{O\,{\scriptsize II}}

\newcommand{\SII}{S\,{\scriptsize II}}
\newcommand{\NII}{N\,{\scriptsize II}}
\newcommand{\NIII}{N\,{\scriptsize III}}

\newcommand{\hst}{\textit{HST}}
\newcommand{\jwst}{\textit{JWST}}
\newcommand{\spitzer}{\textit{Spitzer}}

\graphicspath{{./}{figures/}}

\shorttitle{Improving $z\sim7-11$ galaxy properties with \jwst/NIRCam Medium-Bands}
\shortauthors{Roberts-Borsani et al.}

\begin{document}


\title{Improving $z\sim7-11$ Galaxy Property Estimates with \jwst/NIRCam Medium-Band Photometry}



\correspondingauthor{Guido Roberts-Borsani}
\email{guidorb@astro.ucla.edu}

\author[0000-0002-4140-1367]{Guido Roberts-Borsani}
\affiliation{Department of Physics and Astronomy, University of California, Los Angeles, 430 Portola Plaza, Los Angeles, CA 90095, USA}

\author[0000-0002-8460-0390]{Tommaso Treu}
\affiliation{Department of Physics and Astronomy, University of California, Los Angeles, 430 Portola Plaza, Los Angeles, CA 90095, USA}

\author[0000-0002-3407-1785]{Charlotte Mason}
\affiliation{Center for Astrophysics, Harvard \& Smithsonian, 60 Garden St, Cambridge, MA 02138, USA}
\affiliation{Hubble Fellow}

\author[0000-0002-3418-7251]{Kasper B. Schmidt}
\affiliation{Leibniz-Institut f\"{u}r Astrophysik Potsdam (AIP), An der Sternwarte 16, D-14482 Potsdam, Germany}

\author[0000-0001-5860-3419]{Tucker Jones}
\affiliation{Department of Physics, University of California, Davis, 1 Shields Avenue, Davis, CA 95616, USA}

\author[0000-0003-3820-2823]{Adriano Fontana}
\affiliation{INAF – Osservatorio Astronomico di Roma, Via Frascati 33, I-00078 Monte Porzio Catone (RM), Italy}

\begin{abstract}
The past decade has seen impressive progress in the detection of $z>7$ galaxies with the \textit{Hubble Space Telescope}, however little is known about their properties. The \textit{James Webb Space Telescope} will revolutionise the high-$z$ field by providing NIR (i.e., rest-frame optical) data of unprecedented depth and spatial resolution. Measuring galaxy quantities such as resolved stellar ages or gas metallicity gradients traditionally requires spectroscopy, as broad-band imaging filters are generally too coarse to fully isolate diagnostics such as the 4000 \AA\ (rest-frame) break, continuum emission from aged stars, and key emission lines (e.g., [\OII], [\OIII], H$\beta$). However, in this paper, we show that adding NIRCam images through a strategically chosen medium-band filter to common wide-band filters sets adopted by ERS and GTO programs delivers tighter constraints on these galactic properties. To constrain the choice of filter, we perform a systematic investigation of which combinations of wide-band filters from ERS and GTO programs and single medium-band filters offer the tightest constraints on several galaxy properties at redshifts $z\sim7-11$. We employ the JAGUAR extragalactic catalogs to construct statistical samples of physically-motivated mock photometry and conduct SED-fitting procedures to evaluate the accuracy of galaxy property (and photo-$z$) recovery with a simple star-formation history model. We find that adding $>4.1 \mu$m medium filters at comparable depth to the broad-band filters can significantly improve photo-$z$s and yield close to order-of-magnitude improvements in the determination of quantities such as stellar ages, metallicities, SF-related quantities and emission line fluxes at $z\sim8$. For resolved sources, the proposed approach enables spatially-resolved determination of these quantities that would be prohibitive with slit spectroscopy.
\end{abstract}


\section{Introduction}
One of the major endeavours of modern observational cosmology is to paint a coherent picture of the history of the Universe. To this end, the final frontier remains the identification and characterisation of the first sources that appeared in the Universe, those which played a significant role in reionising the intergalactic medium from a neutral state to a fully ionised one over the first billion years (corresponding to redshifts of $6\lesssim z\lesssim12$). Extragalactic surveys (of deep fields as well as lensing clusters; \citealt{grogin11,koekemoer11,bradley12,ellis13,bradley14,schmidt14,treu15,lotz17,salmon18,coe19}) with the Hubble Space Telescope (\hst) have yielded impressive gains in the number of galaxy candidates at redshifts $z=7-10$, with samples reaching over 1000 objects, and revolutionised our understanding of galaxy evolution therein. Complementing these observations, the spectroscopic confirmation (e.g., \citealt{finkelstein13,oesch15,zitrin15,rb16,hoag17,stark17,hashimoto18}) and characterisation (e.g., \citealt{laporte17a,mainali18,endsley20}) of over a dozen sources has seen impressive advances with ground-based spectroscopy (e.g., probing the rest-frame UV and FIR with Keck/MOSFIRE, VLT/X-Shooter and ALMA), particularly for the brightest and rarest objects. For the rest-frame optical, however, the \textit{Spitzer Space Telescope} has, until now, afforded the only realistic means for statistical analyses. However, the Infrared Array Camera's (IRAC) coarse spatial resolution and the limited depth probed by many surveys makes robust and uncontaminated constraints on galaxy properties a challenging feat. Further advances with current facilities are challenging owing to the limited wavelength coverage of \hst\ and the observed faintness of star-forming galaxies as one approaches redshifts of $z>10$. The imminent arrival of the \textit{James Webb Space Telescope} (\jwst) has the potential to detect galaxies well beyond the current frontier of $z\sim12$ (e.g., \citealt{behroozi20}) thanks to the unprecedented resolution and sensitivity of its NIR imaging and spectroscopic capabilities, and revolutionise our current understanding of galaxy evolution.

The first observations with the observatory will be carried out through accepted ERS and GTO programs and will showcase each instrument's working capabilities as well as set the benchmark for future science with Cycle 1 and beyond. While the ERS and GTO data sets on their own will undoubtedly be extremely valuable for the community (see photometric parameter recovery analyses by e.g., \citealt{kemp19} and \citealt{kauffmann20}), the short life-span of \jwst\ makes extracting all possible data with the highest accuracy imperative, in order to build on the progress achieved until now. Most such observations will carry out wide-band imaging spanning wavelength ranges of $\sim$0.9-4.4 microns, with observations at the reddest end of that range proving particularly valuable for the characterisation of the rest-frame optical of $z\gtrsim7$ galaxies, something previously illustrated by \spitzer/IRAC photometry (e.g., \citealt{labbe13,oesch14,smit15,rb16,castellano17}). However, such observations lack the wavelength resolution to make precise measurements of galaxy properties linked to the state of the gas and underlying stellar populations (see e.g., \citealt{labbe13,rb20}). Even with \jwst, spectroscopy of a large number of sources is expensive and cannot reach the faint limits of the photometric catalogs. 

In this paper, we demonstrate that medium-band photometry provides a cost-effective complement to spectroscopy, allowing for the determination of important physical parameters of galaxies for large and faint samples of galaxies. We reach this conclusion by carrying out a systematic study of the medium-band NIRCam filters and the gain in accuracy that each affords. We find that the addition of a single $>4.1 \mu$m filter yields precise photo-$z$s and galaxy properties from the rest-frame optical for star-forming galaxies at $z\sim7-11$. The paper is structured as follows: we describe the relevant \jwst/NIRCam data sets and construction of mock catalogs in \S2, the SED-fitting techniques and their results in \S3 and our conclusions in \S4. Where relevant, we assume \textit{H}$_{0}=$70 km/s/Mpc, $\Omega_{m}=$0.3, and $\Omega_{\wedge}=$0.7. All magnitudes are in the AB system \citep{oke83}.

\section{JWST programs \& galaxy catalogs}
\subsection{ERS/GTO programs and chosen filters}
\label{subsec:programs}
The main goal of the present paper is to determine the combinations of \jwst/NIRCam medium-band filters which, in addition to observations with wide-band filters yield the most precise recovery of galaxy properties from the rest-frame optical at high redshift. We take as a baseline ERS and GTO programs, which are likely to yield the first large samples of new galaxy candidates due to their sky coverage and unprecedented depths in the NIR with NIRCam imaging, however our conclusions apply to any similar collection of wide-band filters. In constructing our fiducial broad-band setup, the following extragalactic programs are of greatest interest to this analysis (due to the area of the sky probed): ERS 1324 (``\textit{Through the Looking GLASS: A JWST Exploration of Galaxy Formation and Evolution from Cosmic Dawn to Present Day}'', PI: Treu), ERS 1345 (``\textit{The Cosmic Evolution Early Release Science}'', PI: Finkelstein), GTO 1176 (``\textit{JWST Medium-Deep Fields}'', PI: Windhorst), GTO 1199 (``\textit{The Metallicity of Galaxies in the MAC J1149.5+2223 Field}'', PI: Stiavelli), GTO 1208 (``\textit{The CAnadian NIRISS Unbiased Cluster Survey}'', PI: Willott) and GTO 1180/1181 (``\textit{NIRCam-NIRSpec Galaxy Assembly Survey - GOODS-S/N}'', PI: Eisenstein). The primary objectives and instruments used in each program vary and are beyond the scope of this paper, however their common denominator are medium-to-deep ($m\sim$26-30 AB) NIRCam imaging over either well-known cluster/blank fields or new patches of the sky with parallel observations. Of the aforementioned programs, all will carry out observations with the full suite of available wide-band filters from 0.9-4.4 microns (the only exceptions being no F090W imaging for ERS 1345, F277W imaging for GTO 1176, F115W imaging for GTO 1199, and F200W and F356W imaging for GTO 1208), while some will also carry out a modest amount of long-wavelength, medium-band imaging (all programs but ERS 1324 and GTO 1199 will carry out additional F410M imaging, while GTO 1180/1181 will also have F335M observations and GTO 1208 will also have F335M and F360M imaging). While each program will have differing depths to suite their science goals, the significant overlap in filters used between programs means our analysis and results are largely applicable to each of these programs. We thus use as our fiducial set of wide-band filters those from the ERS 1324 program - namely F090W, F115W, F150W, F200W, F277W, F356W, and F444W - and adopt their limiting (5$\sigma$) depth of $\sim$29.4 AB. Additionally, for each of the medium-band filters added to the observations, we adopt the same depth. Finally, we note that although many of the programs will be complemented by existing deep \hst\ and \spitzer\ photometry, here we focus only on the \jwst\ observations, since they are expected to supersede in both depth and resolution all prior data for galaxies in the $z\sim7-11$ redshift range. For simplicity, much of our discussion refers to integrated galaxy photometry, however we emphasise that given sufficient angular resolution and signal-to-noise our results also apply to spatially-resolved properties and thus the strategies discussed here are valid for both types of analyses.

\subsection{JWST/NIRCam colours of high-z galaxies}
\label{subsec:colors}
The rest-frame optical portion of a galaxy spectrum (or SED) offers a plethora of useful features with which to characterise the state of the underlying stellar populations and gas. Chief among these are nebular emission lines from the Balmer series (e.g., H$\alpha$, H$\beta$ and H$\delta$) and from ionic species of Oxygen (e.g., [\OII]$\lambda\lambda$3726,3729 \AA\ and [\OIII]$\lambda\lambda$4959,5007 \AA), Nitrogen (e.g., [\NII]$\lambda$6583 \AA) and Sulphur (e.g., [\SII]$\lambda\lambda$6717,6731 \AA), whose relative ratios and absolute fluxes provide constraints on the dominant rate of star formation from young stars, the gas-phase metallicity, gas electron temperature and density, dust contents and also AGN activity (i.e., the ionisation source of the gas). While nebular emission lines longward of [\OIII]$\lambda$5007 \AA\ are unobservable at $z>7$ with NIRCam, lines blueward of this are (we note that at $z\gtrsim9$ the [\OIII]$\lambda$5007 \AA\ line also drops out of the NIRCam filters). The primary contributor to continuum levels at rest-frame optical wavelengths is continuum emission from aged stars, thus accurate fits of the continuum also provide important estimates of the dominant fraction of a galaxy's stellar mass. Combining this with measurements of the so-called ``Balmer break'' and H$\delta$ emission, these diagnostics can reveal the age of the underlying stellar population (particularly stars with ages $>$10 Myrs) and the timing of the most recent burst of star formation. Such diagnostics are far fewer in the rest-frame UV (e.g., a lack of strong emission lines), which is in general well-sampled by \hst/WFC3 and will be further by upcoming NIRCam wide-band observations.

\begin{figure}
\center
 \includegraphics[width=\columnwidth]{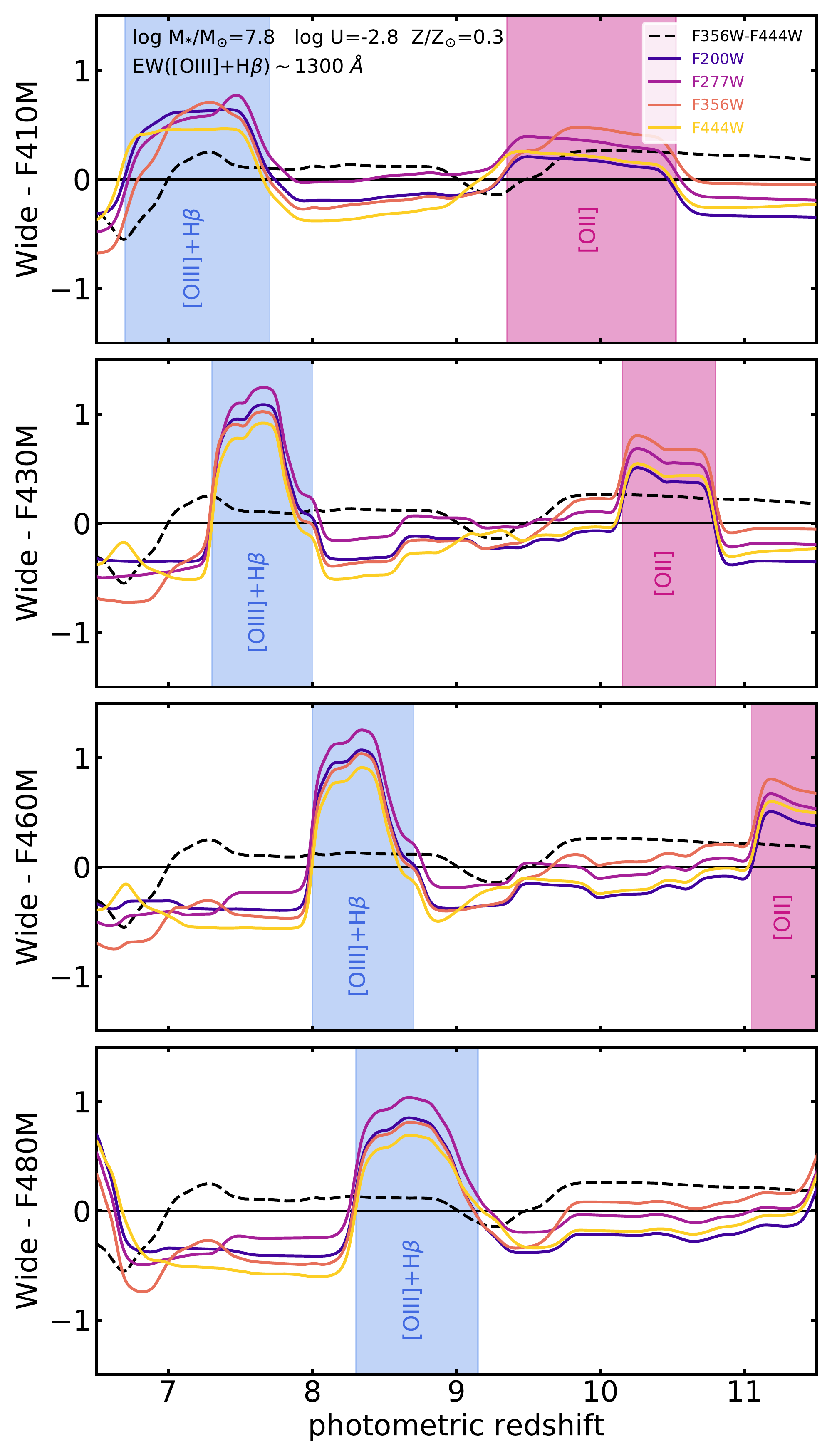}
 \caption{The \jwst/NIRCam wide- minus medium-band colours as a function of photometric redshift for the (from top to bottom) F410M, F430M, F460M and F480M filters, where [\OIII]+H$\beta$ is likely to transit at $z>7$. Additionally, the F356W$-$F444W colours are also plotted as a dashed black line. The blue and magenta fill in each panel highlights the region where particularly red colours - driven by [\OIII]+H$\beta$ and a Balmer break plus [\OII]+[\NIII]+H$\delta$, respectively - isolate a narrow photometric redshift range, providing a useful constraint in the absence of deep optical imaging where the Lyman-break is likely to pass through.}
 \label{fig:IRcolors}
\end{figure}

Each of these diagnostics contribute to and influence the colours of galaxies at IR wavelengths (to varying degrees), which can be useful in identifying galaxies with e.g., particularly low or high levels of star formation or that are likely to lie within a particular photometric redshift range. For this latter consideration, several studies have demonstrated the effectiveness of using IR colour cuts with deep \spitzer/IRAC imaging to constrain the photometric redshifts of high-$z$ galaxies in the absence of deep optical bands (see e.g., \citealt{labbe13,smit15,rb16}). The contamination of the 3.6 $\mu$m channel by H$\alpha$ emission and later strong [\OIII]+H$\beta$ emission in the 4.5 $\mu$m channel produces particularly blue [3.6]$-$[4.5] colours at $z\sim6.8$ \citep{smit15} and particularly red [3.6]$-$[4.5] colours at $z\sim7-9$ \citep{rb16}, respectively. Due to the large bandwidth of the aforementioned IRAC channels ($\sim$7000 \AA\ and $\sim$8500 \AA, respectively), however, the isolation of [\OIII]+H$\beta$ at $z>7$ is challenging and thus the cause of the especially red colours across the $z\sim7-9$ is not always unambiguous, thanks to the nearby presence of the rest-frame 4000 \AA\ and Balmer breaks (see \citealt{rb20}). This makes a more precise photometric redshift selection across the $z\gtrsim7-9$ range with IR colours challenging. Fortunately, the arrival of \jwst/NIRCam medium-band filters offers useful alternatives: the reduced bandwidth of the medium-band filters offer a cleaner measurement of nebular emission lines and continuum emission, thereby removing (in part) several degenerate solutions. In Figure \ref{fig:IRcolors} we plot the NIRCam colours between our fiducial set of wide-band filters (from F200W longward, since these are unaffected by the Lyman-break until $z\gtrsim$13) and each of the medium-band filters which could be contaminated by [\OIII]+H$\beta$ at $z>7$ (specifically, F410M, F430M, F460M, and F480M) for a young (20 Myr) galaxy SED (the exact parameters are labelled in the plot), in order to discern which photometric redshift ranges would be particularly well constrained by red IR colours. For reference, we also plot F356W$-$F444W colours since these are virtually identical to the first two \spitzer/IRAC bands that are traditionally used for this exercise. We find four redshift ranges not seen in wide-band colours where especially red medium-band colours could significantly aid in constraining the photo-$z$. For each of the F410M, F430M, F460M and F480M filters, these ranges peak approximately at $z\sim7.2$, $z\sim7.7$, $z\sim8.4$ and $z\sim8.7$ and have narrow ranges $\Delta z\approx1.0$, $\Delta z\approx0.7$, $\Delta z\approx0.7$ and $\Delta z\approx0.85$, respectively. However, we do not observe any major differences across the majority of our photo-$z$ range with the choice of the wide-band filter as long as it remains unaffected by the Lyman-break and rest-frame optical. Within the $z=7-12$ range, we do however note far smaller and less constrained trends of red colours at redshifts of $z\sim10.4$ and $z\sim11.25$ using the F430M and F460M filters, respectively. These are due to the combination of a Balmer break and/or the presence of less strong emission lines such as [\OII]$\lambda\lambda$3726,3729 \AA, [\NIII]$\lambda\lambda$3868,3967 \AA\ and H$\delta$. Of course, the amplitudes of the curves are highly dependent on the EW of the nebular emission lines (in particular [\OIII]+H$\beta$) and increasing them merely increases each of the aforementioned peaks. While such an exercise is typically conducted with line strengths of EW([\OIII]+H$\beta$)$>$2000 \AA\ characteristic of extreme objects, our fiducial model here has a measured EW([\OIII]+H$\beta$)$\sim$1300 \AA, characteristic of less extreme and perhaps more typical objects (see e.g., \citealt{tang19} and \citealt{tang20}): our adopted line strength falls well within the distributions of EWs (typically several hundreds to $\sim$3000 \AA) inferred for bright, $z\sim7-8$ galaxies through \spitzer/IRAC imaging \citep{labbe13,rb16,debarros19,endsley20}. It is thus extremely encouraging that even less extreme and more representative galaxies are able to produce much more pronounced ($\sim$0.5-1 mag redder) and distinct colour peaks than seen with wide-bad imaging, clearly illustrating the added benefit of using medium-band filters (in combination with appropriate imaging blueward of the Lyman break to support dropout techniques) to securely pre-select galaxies at photometric redshifts of $z\sim7.2$, $z\sim7.7$, $z\sim8.4$, $z\sim8.7$ (and also $z\sim10.4$ and $z\sim11.25$), compared to using their wide-band counterparts only. This is particularly relevant for observations lacking deep $\sim$1 $\mu$m observations (but still including sufficient imaging for dropout selections) with which to constrain the location of the Lyman-break to high precision. NIRCam's medium-band filters provide an unprecedented opportunity to better sample photometric redshifts and galaxy properties through cleaner measurements of emission lines and continua over rest-frame optical wavelengths at $z>7$.

\subsection{Mock catalogs of star-forming galaxies}
To determine and evaluate the uncertainties on rest-frame optical galaxy properties (global and resolved) that the programs described in Section \ref{subsec:programs} are likely to yield, a statistical approach is required, where photometric catalogs of representative galaxies are fitted with appropriate SED-fitting techniques over the filter combinations of interest. Thus, we begin by constructing distributions of representative global galaxy and star-formation history properties with which to randomly derive input parameters for the generation of mock SEDs and photometry at high redshift.
 
 \begin{figure*}
\center
 \includegraphics[width=1.75\columnwidth]{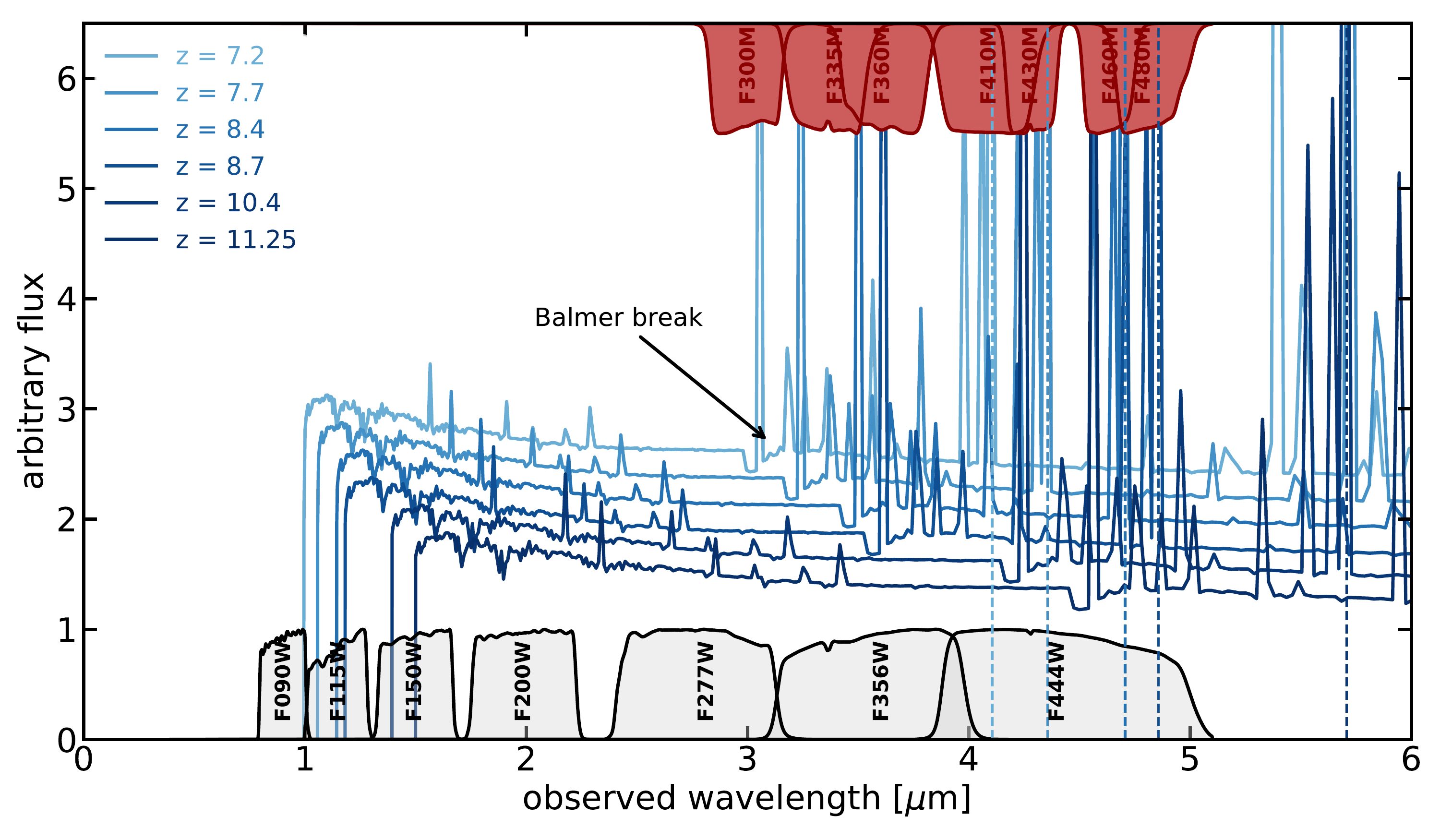}
 \caption{A mock star-forming galaxy SED at our fiducial $z>7$ redshift intervals and the corresponding wide-band \jwst/NIRCam filter response curves (bottom, gray) and red medium-band NIRCam filters (red, top) used throughout this study. The dashed vertical line marks the position of the [\OIII]$\lambda$5007 \AA\ emission line at the various redshifts, while the position of the Balmer break at $z=7.2$ is also indicated.}
 \label{fig:sedfilters}
\end{figure*}

\begin{figure*}
\center
 \includegraphics[width=2.\columnwidth]{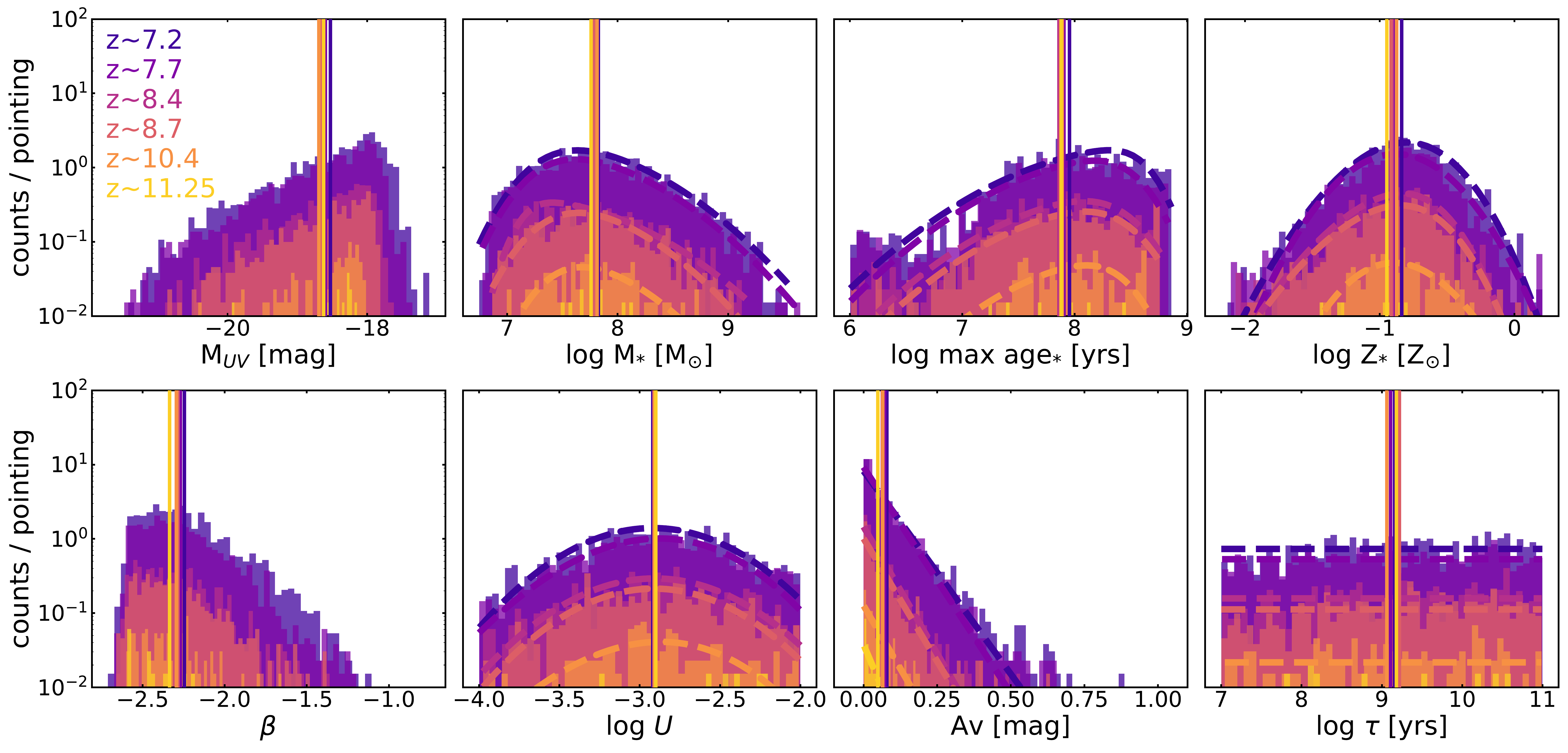}
 \caption{The distribution of integrated galaxy properties from the JAGUAR catalogs, extracted for all star-forming galaxies at redshifts $z\sim7.2$, $z\sim7.7$, $z\sim8.4$, $z\sim8.7$, $z\sim10.4$ and $z\sim11.2$ and associated skew-normal fits (dashed lines of the same colours). Properties without a dashed line are not used as input to \texttt{Bagpipes}. The vertical lines denote the mean value of the distribution at each redshift interval and each histogram is normalised by the combined pointing area ($\sim$9.25 arcmin$^{2}$) of both NIRCam modules (A and B) for red wavelengths, to illustrate the expected number of sources per pointing.}
 \label{fig:props}
\end{figure*}

The galaxy properties are derived from each of the 10 realisations of the v1.2 JAdes extraGalactic Ultradeep Artificial Realizations (JAGUAR; \citealt{williams18}) mock catalogs, which make use of an empirical model driven entirely by observational trends at $0.2<z<10$ to create mock observations of galaxies out to $z=15$. Each realisation covers an area of 121 arcmin$^{2}$, thus the total area probed by the combined samples spans 1210 arcmin$^{2}$. Since rest-frame optical wavelengths remain inaccessible with even the reddest medium-band NIRCam filters at $z>12$, we opt to carry out our analysis at a slightly lower redshift interval of $7\lesssim z\lesssim12$, where such wavelengths are still accessible. Specifically, given the improved photometric constraints with the F410M, F430M, F460M and F480M filters discussed in Section \ref{subsec:colors}, we opt for redshift intervals of $z=7.2$, $z=7.7$, $z=8.4$, $z=8.7$, $z=10.4$ and $z=11.25$. For illustration, the star-forming SED used in Section \ref{subsec:colors} is shown in Figure \ref{fig:sedfilters} at the various redshift intervals, along with the NIRCam filters of interest (i.e., all of our fiducial wide-band filters and the red, medium-band filters). Additionally, since even the deepest ERS and GTO programs will reach depths up to m$\sim29$ AB, we focus on galaxies with m$_{F200W}\sim$26-29 AB. In Figure \ref{fig:props} we show the distribution of global galaxy properties (centered on our fiducial redshift intervals and over the aforementioned magnitude limits) of the JAGUAR mock galaxy samples described above, which we use as the basis to extract mock photometry in this study. We allow the distributions to span $\Delta z=$0.25 either side of our fiducial redshifts, except for $z=8.4$ and $z=8.7$ where we reduce this to $\Delta z=$0.15 in order to prevent overlap within the redshift ranges. Additionally, we also consider sources with log $U < -$2 in order to consider ``normal'' galaxies of the early Universe.

To construct our mock photometry, we use the \texttt{Bagpipes} \citep{carnall18} SED-fitting code, which is fed randomly drawn properties from the JAGUAR distributions. We opt for this method rather than using the JAGUAR photometry itself, since the number density of high-$z$ sources in JAGUAR follows observed trends and thus drops dramatically as a function of redshift, making a statistical analysis challenging for our purposes. More specifically, we use as input properties the stellar mass (M$_{*}$), maximum stellar age, metallicity ($Z$, assumed to be the same for both the stars and the gas), ionisation parameter for nebular emission (log $U$), dust attentuation (A$_{\text{v}}$; assuming a \citealt{calzetti2000} attenuation law) and the star formation timescale ($\tau$) for a declining star formation history model. The velocity dispersion of absorption and emission lines is fixed at 150 \kms. To ensure the derived photometry is representative of the general galaxy population, we fit each of the relevant JAGUAR distributions with a skew-normal (log M$_{*}$, stellar age, $Z$ and log $U$), uniform ($\tau$), or exponential (A$_{\text{v}}$) profile and sample their resulting PDFs to generate the random values. Since stellar mass and stellar age show some minor correlation, these are sampled jointly. We note here that since the distributions are quite similar across our redshift bins, they are unlikely to introduce strong biases in our analyses. 100 different mock galaxies SEDs are subsequently generated from the randomly drawn parameters and the photometry is then extracted for the wide-band filters and all red, medium-band NIRCam filter. The procedure is repeated for each redshift bin and F200W apparent magnitudes of m$_{F200W}\sim$26-29 AB for $z=7.2-8.4$ galaxies, m$_{F200W}\sim$27-29 AB for $z=8.7$ galaxies, m$_{F200W}\sim$28-29 AB for $z=10.4$ galaxies and  m$_{F200W}\sim$29 AB for $z\sim11.25$ galaxies. The magnitude bin centers are located at each integer value of the considered range, in steps of 1 mag and widths of $\pm$0.5 mag. The associated 1$\sigma$ errors are derived by scaling and combining the 5$\sigma$ ERS 1324 depth to an additional 2\% absolute calibration\footnote{\url{https://jwst-docs.stsci.edu/data-processing-and-calibration-files/absolute-flux-calibration}} uncertainty, for each filter. In total, accounting for the redshift and magnitude bins the above results in 1,800 mock galaxy SEDs from which to extract NIRCam photometry.

\section{SED-fitting and Results}
\subsection{The accuracy of galaxy properties}

\begin{table}
    \centering
    \begin{tabular}{lc}
        \hline
        Parameter & Allowed Range \\
        \hline
        Redshift & [$0,15$]\\
        Mass formed & [$6,10$] M$_{\odot}$\\
        Maximum stellar age & [$0.001,1$] Gyr\\
        Metallicity & [$0-5$] Z$_{\odot}$\\
        log $U$ & [$-4,-2$] \\
        A$_{\text{v}}$ & [$0,3$] mag \\
        $\tau$ & [$>0.01$] Gyr\\
        \hline 
    \end{tabular}
    \caption{The free parameters in our declining star formation history model and allowed ranges, used to fit our mock NIRCam photometry with \texttt{Bagpipes}.}
    \label{tab:params}
\end{table}

\begin{figure*}
\center
 \includegraphics[width=1.4\columnwidth]{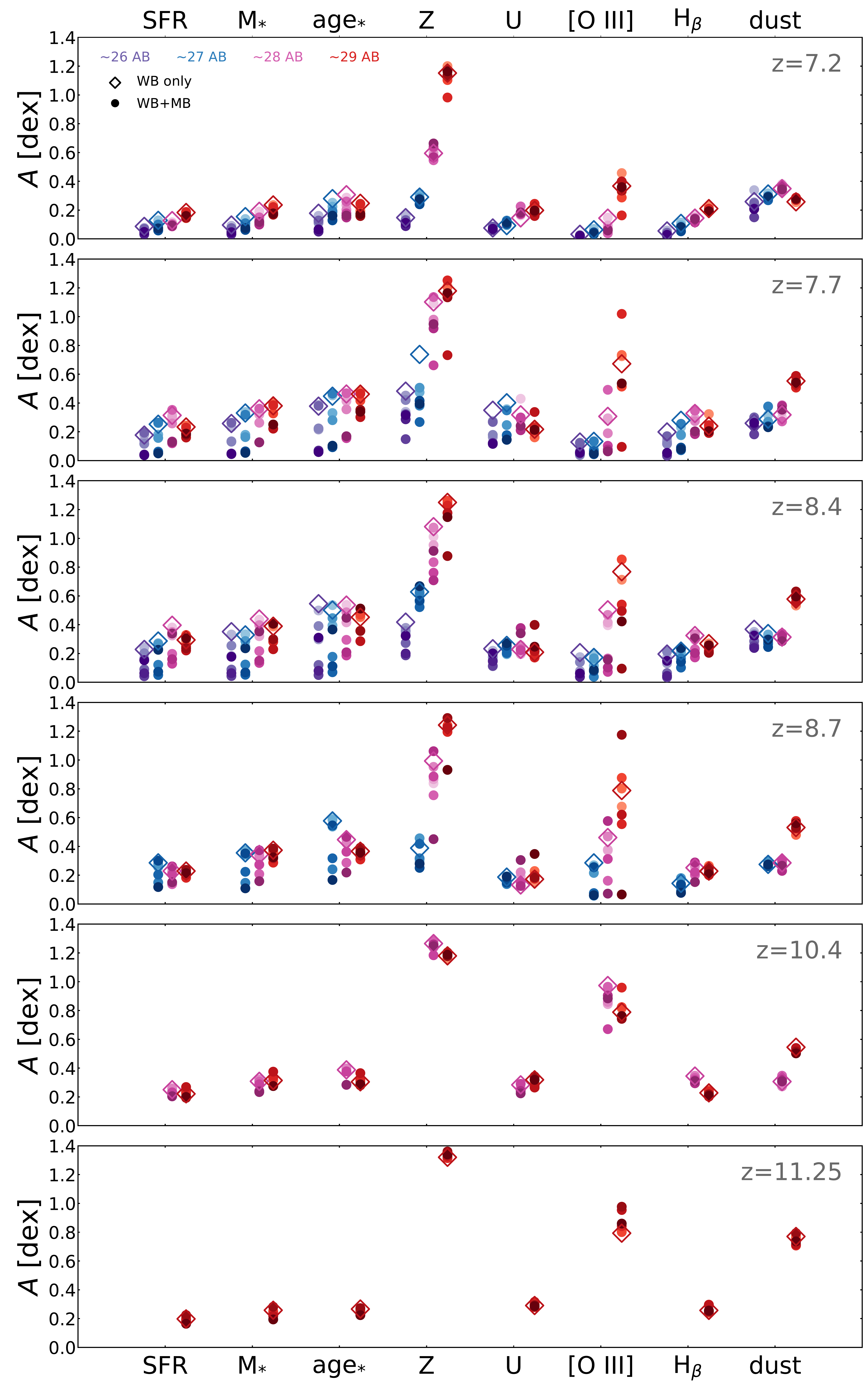}
 \caption{The results of the SED-fitting procedure for \jwst/NIRCam photometry. The $x$-axis denotes a galaxy property/free parameter of interest (from left to right: star formation rate, stellar mass, maximum stellar age, metallicity, ionisation parameter, [\OIII]$\lambda$4959  \AA+[\OIII]$\lambda$5007 \AA\ flux, H$\beta$ flux and dust extinction), while the $y$-axis denotes the median (and absolute) residual between the (log) true and (log) recovered parameter, according to each redshift, magnitude, and filter combination. Diamonds represent fitting results with the wide-band filters only, while each circle denotes a combination of wide filters plus an additional red medium-band filter. The F200W magnitudes of the galaxy models are distinguished by the different colour maps (purple, blue, pink, and red shades), while the shade of each point in a map dictates the medium filter used, with darker shades corresponding to redder wavelengths.}
 \label{fig:results}
\end{figure*}

Each iteration and combination of galaxy photometry is subsequently fit using \texttt{Bagpipes}, this time adopting uniform priors on each of the parameters that were used in the construction of the mock galaxy catalogs, including their redshifts. The allowed ranges are presented in Table \ref{tab:params}. While the observed properties of galaxies are highly dependent on (and also degenerate with) the chosen star formation history, the goal here is not to retrieve the true star formation histories but rather to determine the accuracy of each retrieved parameter as a function of filter combination and input parameter. With this in mind, we present the main results of our fits for representative, star-forming galaxies in Figure \ref{fig:results}, where we plot the accuracy of recovered galaxy properties as a function of wide+medium filter combination. Here we define the accuracy, $A$, as the median (over the 100 iterations) absolute difference between the input and recovered parameter:

\begin{equation}
A = |\,\text{log}_{10}(\text{param}_{\text{true}}) - \text{log}_{10}(\text{param}_{\text{fit}})\,|
\end{equation}

We begin by exploring the accuracy of recovered properties such as galaxy SFR, stellar mass (M$_{*}$), maximum stellar age, metallicity, ionisation parameter ($U$), emission line fluxes (specifically [\OIII]$\lambda\lambda$4959,5007 \AA\ and H$\beta$), and dust attenuation. Not all of these are independent quantities and will have some correlation - the SFR and related quantities are calculated as the average of the preferred SFH over the last 100 Myrs. Considering first results using the set of wide-band filters only (marked by the empty diamond symbols), we find the recovery generally fares reasonably well, since either (or both) the F356W or F444W filter probe the rest-frame optical. In particular, measurements of SFR, stellar mass, maximum stellar age, ionisation parameter, H$\beta$ emission and dust contributions generally appear accurate to less than $\sim$0.6 dex at all redshift intervals. Metallicity measurements and [\OIII]$\lambda\lambda$4959,5007 \AA\ line fluxes tend to fare worse than the other parameters, with accuracy levels spanning values of $\sim$0.2-1.2 dex for both, depending on the redshift interval. We note that the accuracy for brighter galaxies fares on average better ($\sim$0.2-0.6 dex) than that of fainter galaxies ($\sim$0.6-1.2 dex) - this is not unexpected, since the higher signal-to-noise of the photometry is likely to yield more accurate constraints.

In virtually all cases and at all redshifts, the addition of a well-chosen medium-band filter improves the accuracy, due to less contaminated measurements of the continuum or key emission lines. This is particularly evident at redshift intervals less than $z\sim9$, where the rest-frame optical portion of the spectrum is still well-sampled by NIRCam. At $z=10-11$, the gains in accuracy become much more marginal, since the rest-frame optical portion of the spectrum begins to drop out of even the reddest filter (see Figure \ref{fig:sedfilters}). The gains in accuracy are best seen with bright galaxies, given their higher S/N, however the improvement remains significant even for their fainter counterparts. For most properties (e.g., SFR, stellar mass, stellar age), the improvements span $\sim$0.2-0.5 dex. However for others, such as galaxy metallicity and [\OIII] flux, the improvements can be more significant ($>0.5$ dex). Such improvements are particularly important and useful when found for quantities such as the SFR, stellar mass and metallicity, since such quantities are paramount towards characterising the mass-metallicity and ``main sequence'' relations at high-$z$. Typical measurements using \hst\ and \spitzer\ are subject to large ($>$0.5-1 dex) uncertainties which make accurate analysis challenging. Reducing those uncertainties to only $\sim$0.1-0.2 dex is paramount towards gaining an accurate picture of the mass build up and gas and metal cycling of galaxies near cosmic dawn. Here we show that the addition of a single, strategically placed medium-band filter can improve the precision of such quantities significantly, with resulting accuracies of $\sim$0.1-0.4 dex (SFR and stellar mass) and $\sim$0.2-0.3 dex for the metallicity of bright galaxies (which increases to $\sim$0.6-1 dex for faint galaxies).

Further quantifying the gain in accuracy, for a given magnitude ($m$) and redshift bin we define the accuracy gain ($g$) of a galaxy property ($p$) between the fiducial set of wide-band filters (W) and an additional medium-band filter (M) as 

\begin{equation}
g_{p}(m,z) = \frac{A_{\text{p,W}}(m,z)}{A_{\text{p,M}}(m,z)},
\end{equation}

then we can define the total (median) gain for a galaxy in a given redshift bin ($G(z)$, summed over all magnitude bins and properties presented in Figure \ref{fig:results}) as

\begin{equation}
G(z) = \sum_{m}\sum_{p}g_{p}(m,z),
\end{equation}

\begin{figure*}
\center
 \includegraphics[width=1.5\columnwidth]{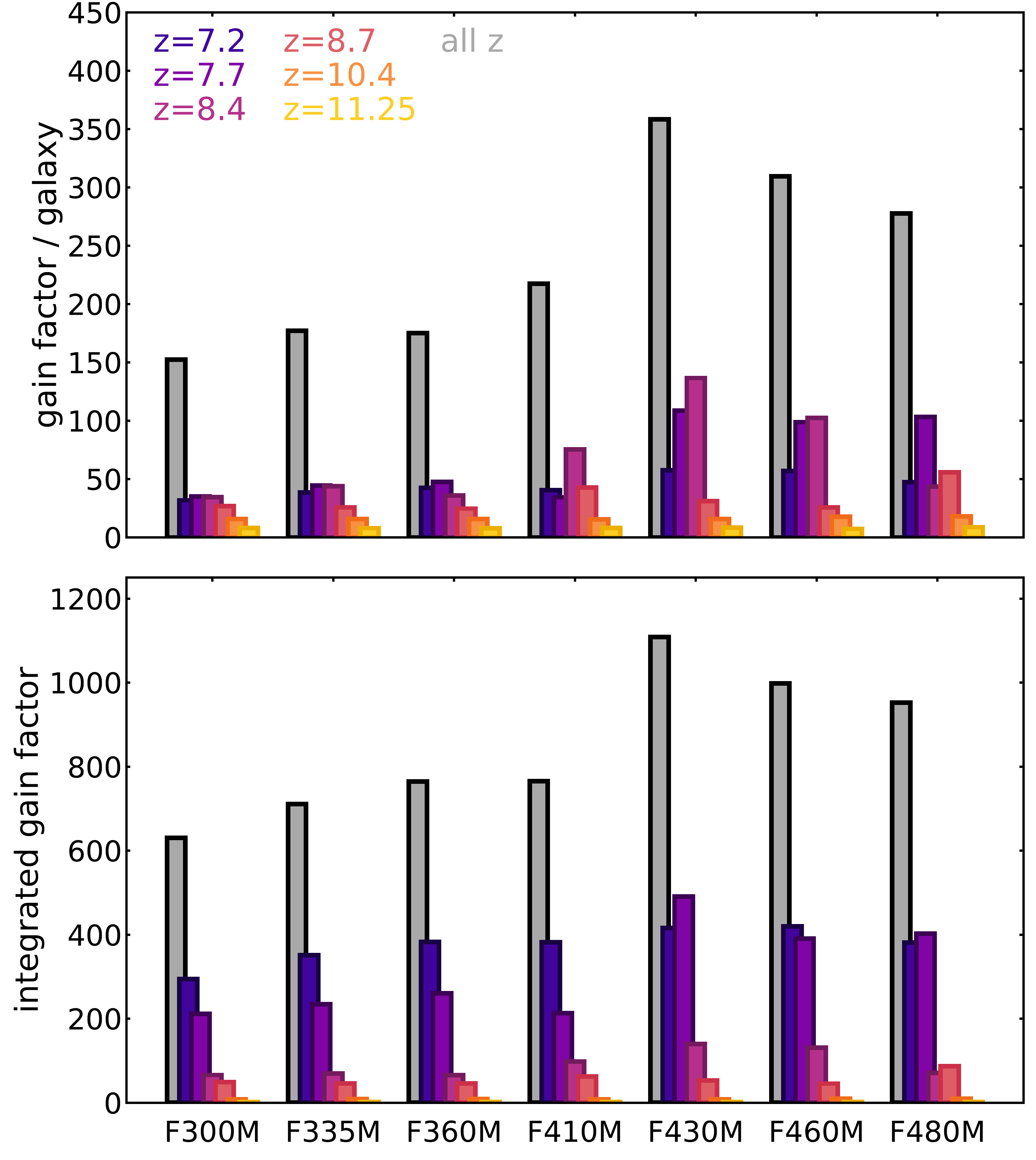}
 \caption{Top: The total accuracy gain (integrated over all properties and magnitudes presented in Figure \ref{fig:results}) per galaxy as a function of redshift and additional long-wavelength MB NIRCam filter, over results inferred with wide-band photometry (adopted by ERS and GTO programs) only. Additionally, the gray bars indicate the summed gains summed over all redshift bins. Bottom: the same as the top but multiplied by the expected number of JAGUAR galaxies per NIRCam pointing in each magnitude and redshift bin. The $\geqslant$4.3 $\mu$m filters clearly provide the highest gain for $z\sim8$ (i.e., $z_{\rm input}=7.7$ and $z_{\rm input}=8.4$) galaxies and overall, while the F430M filter provides the largest gain out of all medium filters. Each redshift bin is colour-coded according to the colour scheme in Figure \ref{fig:props}.}
 \label{fig:gainsummary}
\end{figure*}

We display this gain factor as an integrated quantity per galaxy (i.e., the median value per galaxy summed across all properties and over all magnitude bins) and as a function of redshift and medium-band filter in the top panel of Figure \ref{fig:gainsummary}. We find virtually all medium filters return similar gains for the $z=7.2$ populations (with gain factors of $\times$32-58), while more significant gains are returned for the $z=7.7$ and $z=8.4$ (i.e., $z\sim8$; with gain factors of $\times$44-137) populations using $\geqslant4.3 \mu$m bands: at $z=7.2$ the optical spectrum is already relatively well sampled by the wide-band filters, while at $z=8.7$ and above the rest-frame optical begins to drop out of the filters. More specifically, the large gains for the $z\sim8$ populations seen in Figure \ref{fig:gainsummary} is primarily due to the improved accuracy found for galaxy metallicity (increasing the accuracy by $\sim$0.5 dex) and [\OIII] emission line flux (increasing the accuracy by $\sim0.5-0.8$ dex). We find the F430M filter in particular returns the most significant accuracy gains both for $z\sim8$ galaxies ($z_{\rm input}=[7.7,8.4]$) and overall (i.e., summing the results over all redshift bins) - a factor of $\sim$1.2-3.5 and $\sim$1.15-2.3 better than all other filters, respectively. The most marginal gains are observed at redshifts $z\gtrsim9$, where only a handful of filters probe the bluest regions of the rest-frame optical and the number density of galaxies at such higher redshift drops dramatically; we find no major difference in accuracy gain between any of the medium-band filters at those redshifts, since they all observe similar spectral features (i.e., mostly the rest-frame UV portion of the spectrum, where strong spectral features are lacking). The bottom panel of Figure \ref{fig:gainsummary} highlights the change in gain when accounting for the expected number of JAGUAR galaxies in our fiducial redshift and magnitude bins (i.e., multiplying the top panel results by the number of galaxies that would be observed with a single NIRCam pointing). As expected, the gains at the highest redshifts are reduced compared to their lower-$z$ counterparts, due to the declining density of galaxies as a function of redshift. We note, however, that the F430M filter continues to afford the largest gains for $z\sim8$ galaxies and overall - with factors of $\sim$1.2-2.3 and $\sim$1.1-1.8 better than all other filters, respectively. Thus, considering these results as well as the photo-$z$ improvements from Figure \ref{fig:photoz}, the F430M represents in our view the optimal medium-band filter for maximising accuracy gains in both photo-$z$ and galaxy property estimates. Of course, calculations presented here are for especially deep observations, however significant gains can also be found with shallower data (as illustrated by the fainter galaxies in Figure \ref{fig:results}) which would require shorter observation times.

\subsection{A note on systematic uncertaintes}
\begin{figure*}
\center
 \includegraphics[width=0.75\textwidth]{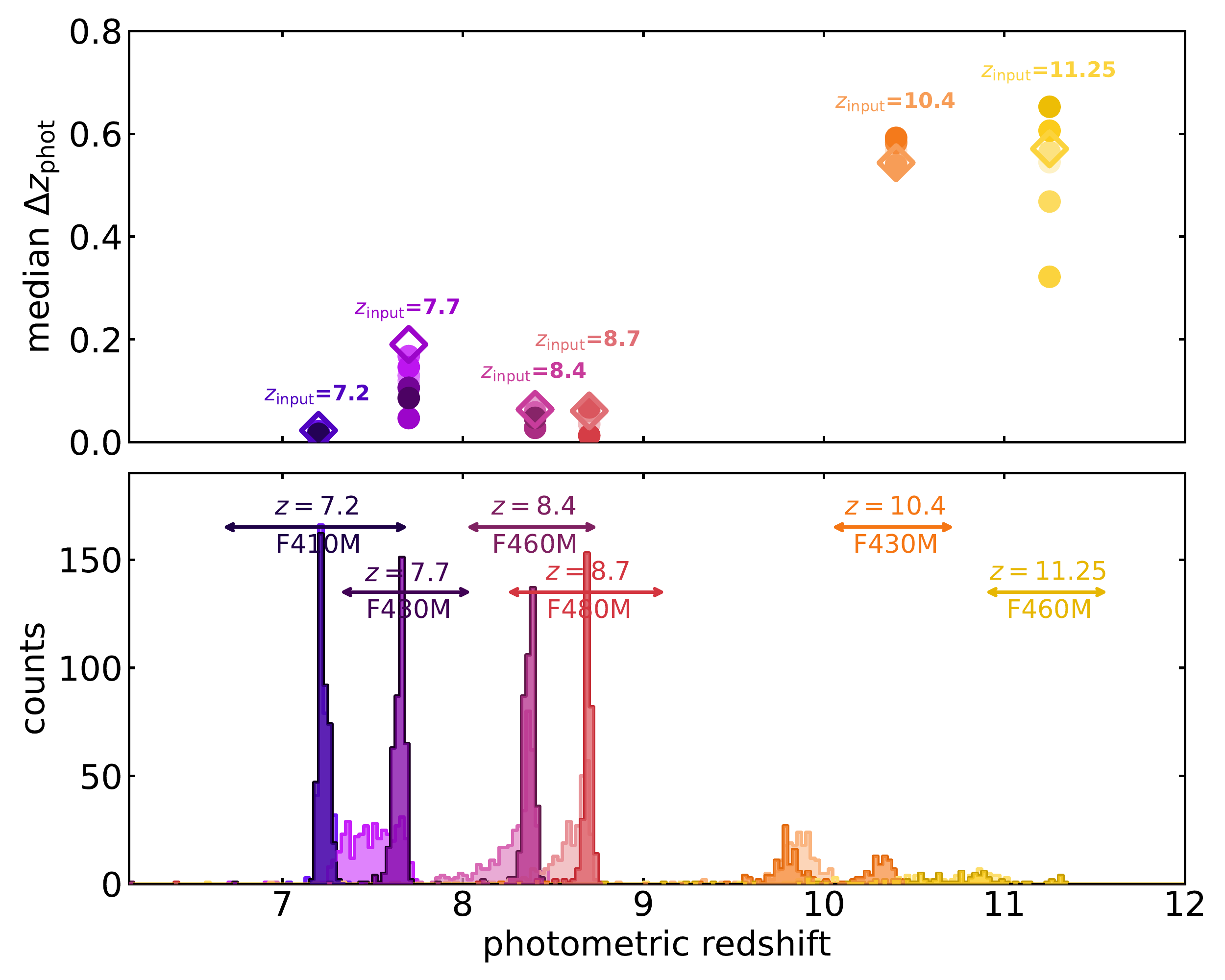}
 \caption{Top panel: The median (absolute) difference between the input redshift and recovered photometric redshift ($|z_{\rm input}-z_{\rm output}|$) as a function of wide-only (empty diamonds) and wide+medium (filled circles) filter combinations. The median is taken over all of the galaxies in a given redshift bin. The redshift colour scheme follows that of Figure \ref{fig:props}, with darker shades of a given colour map indicating redder-wavelength medium filters. Bottom panel: A comparison of the distributions of recovered redshifts using wide filters only (light histograms) or wide+medium filters (darker histograms; F410M for $z\sim7.2$, F430M for $z\sim7.7$ and $z\sim10.4$, F460M for $z\sim8.4$ and $z\sim11.25$, and F480M for $z\sim8.7$; see Figure \ref{fig:IRcolors} and Section \ref{subsec:colors} for details). Clearly, the addition of a strategically-placed $>4$ $\mu$m medium-band filter significantly improves the photo-$z$ estimate at $z<10$. The redshift colour scheme follows that of Figure \ref{fig:props}.}
 \label{fig:photoz}
\end{figure*}

We have performed the relatively simple exercise of parameter recovery adopting SED-fitting of NIRCam photometry and a certain SFH history model, and assessed the precision of the resulting best fit properties. However, such a feat glosses over important caveats that must be taken into consideration. The first is that we have assumed a declining SFH history model. The motivation for this was to be consistent with the models employed by the JAGUAR catalogs, from which we derive our input galaxy property distributions. However, the star formation histories of high-$z$ galaxies are unconstrained and the adoption of a SFH model naturally introduces significant uncertainties which any fitting of photometric data sets is likely to be limited to. Secondly, we have allowed the redshifts of our fitted galaxies to vary, meaning each photometric measurement will have some inherent uncertainty due to the resulting photometric redshift. While in general the redshifts of $z>7$ are already well constrained by deep $Y$- and $J$-band observations, we find the recovered redshifts of $z>7$ are generally better constrained with the addition of a strategically-placed medium-band filter. This is highlighted in the top panel of Figure \ref{fig:photoz}, where we present the (median) absolute difference between input and recovered redshifts over all galaxies in a given redshift bin, in similar fashion to the results presented in Figure \ref{fig:results}. Clearly, the combination of wide-band filters alone returns reasonably well-constrained photo-$z$s, however in virtually every redshift bin the recovered photometric redshift is improved with the addition of a medium-band filter; the maximum difference between photo-$z$s from the various filter combinations are $\Delta z=0.02$ for $z_{\rm input}=7.2$, $\Delta z=0.19$ for $z_{\rm input}=7.7$,  $\Delta z=0.07$ for $z_{\rm input}=8.4$, $\Delta z=0.07$ for $z_{\rm input}=8.7$, $\Delta z=0.59$ for $z_{\rm input}=10.4$ and $\Delta z=0.65$ for $z_{\rm input}=11.25$. For the aforementioned redshifts, although the differences are sometimes small (e.g., $\Delta z=0.02$ for $z_{\rm input}=7.2$), the medium-band filter which returns the most accurate median photo-$z$ is F430M, F430M, F460M, F480M, F430M, and F430M, respectively. The preferred filters for $z_{\rm input}=7.7-10.4$ are consistent with those that straddle strong [\OIII]+H$\beta$ emission lines. Clearly, the F430M filter provides not only the most significant galaxy properties gains, but also samples well-defined spectral features that lead to the best photo-$z$ constraints across the majority of redshift bins.

The improvement of the photo-$z$s is further illustrated in the bottom panel of Figure \ref{fig:photoz}, where we plot the distributions of all recovered photometric redshifts using wide-band filters only (light shaded histograms) and the wide+medium-band combinations (darker shaded histograms) discussed in Section \ref{subsec:colors} (i.e., wide filters plus F410M, F430M, F460M, F480M, F430M and F460M for $z_{\rm input}=7.2$, $z_{\rm input}=7.7$, $z_{\rm input}=8.4$, $z_{\rm input}=8.7$, $z_{\rm input}=10.4$ and $z_{\rm input}=11.25$, respectively), where the medium-band straddles strong [\OIII]+H$\beta$ emission lines. As shown in the top panel of the same figure, we find the latter distributions are peaked far more sharply than the former, are particularly well constrained for $z_{\rm input}<10$ sources, and fall exactly within the expected photo-$z$ range given by the especially red colours discussed in Section \ref{subsec:colors}. The same cannot be said for the $z_{\rm input}=10.4$ and $z_{\rm input}=11.25$ fits, however, due to the sampling of weaker features than the [\OIII]+H$\beta$ lines which result in a less well-defined red colour, and are thus less reliable. The approximate ranges of the red colours (due to [\OIII]+H$\beta$ emission lines at $z<10$ and [\OII]+Balmer break emission at $z>10$) shown in Figure \ref{fig:IRcolors} are also indicated in the bottom panel of Figure \ref{fig:photoz}. Although not shown here, we note that the distributions of photometric redshifts considering \textit{all} the medium-band filters mirrors those of the wide-band filter distributions, thus highlighting in general the constraining power of the Lyman-break but also the increased photo-$z$ accuracy from the isolation of strong nebular emission lines. We conclude that while there are significant uncertainties in any measurement of photometric data sets due to the assumed SFH model, the impact of photo-$z$s on our $z<10$ samples here is not significant and the photometric redshift themselves are generally improved with the addition of a medium filter.

\section{Observational Strategy}

The medium-band filters of greatest interest here are the $>4 \mu$m filters, given the gains in accuracy they afford over the $z\sim7-11$ population (the F300M, F335M and F360M do not probe redward enough to affect the properties of $z\sim9-11$ galaxies), as illustrated in the top panel of Figure \ref{fig:gainsummary}. As discussed in previous sections, while each of these afford similar information gains for $z\sim7.2$ galaxy populations, the latter three afford significantly higher gains in particular for $z\sim8$ galaxies compared to the F410M filter. Pushing to even higher redshifts, only the F460M and F480M probe redward enough to affect $z\sim11$ galaxies, however the number counts of these objects are expected to be comparatively low (see e.g., \citealt{ellis13,oesch18}). Thus, for studies of the $z\sim7.7$ and $z\sim8.4$ populations (i.e., $z\sim8$ galaxies), the F430M filter would be best suited for the most significant accuracy gains in both photo-$z$ and galaxy property estimates.

\begin{figure}
\center
 \includegraphics[width=\columnwidth]{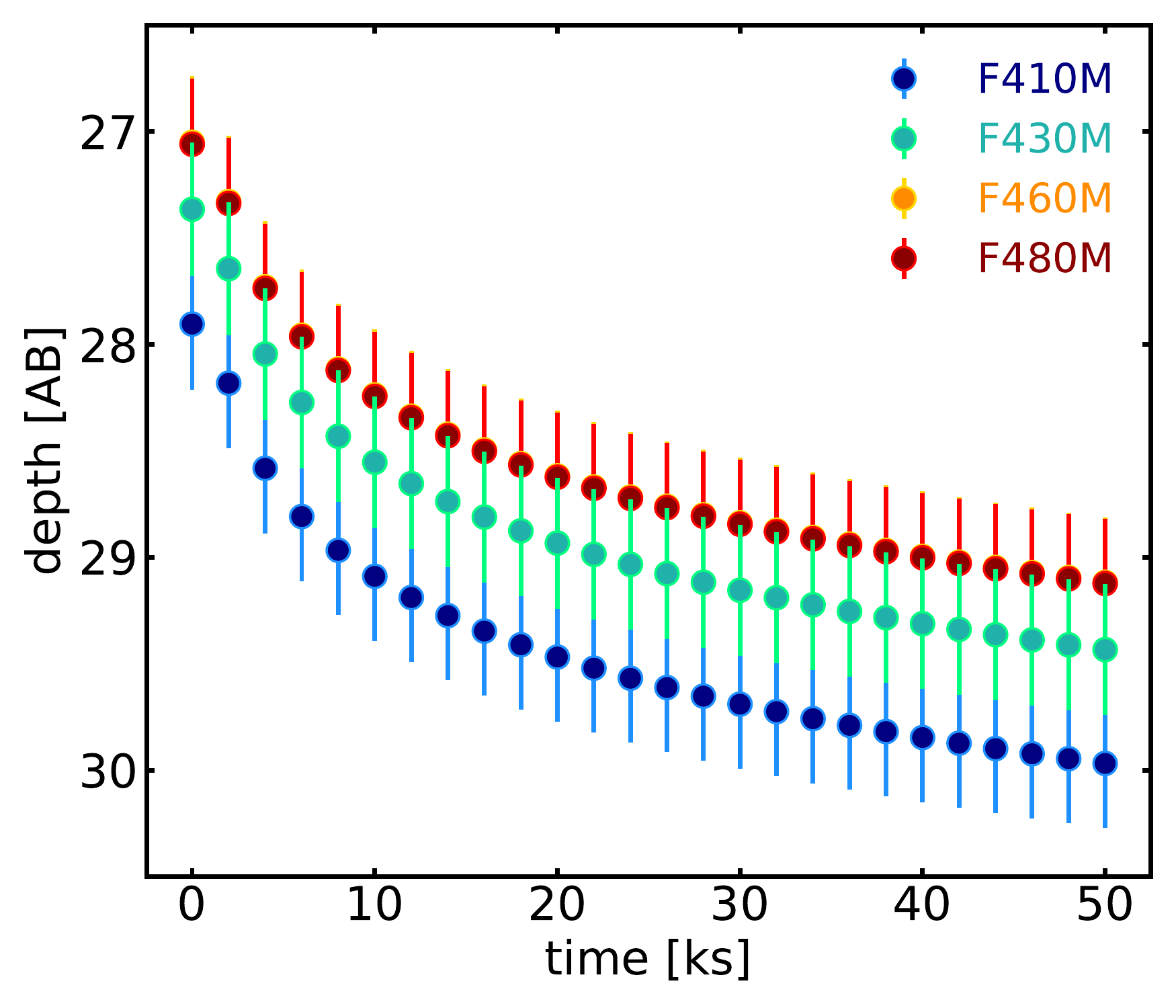}
 \caption{NIRCam imaging (5$\sigma$) depths estimated using the \jwst\ ETC, as a function of time and for a variety of observational settings. Each colour corresponds to a different medium-band filter. Circles mark the median depth between each of the settings, while error bars mark their associated 1$\sigma$ standard deviation. Results for the F460M filter are virtually identical to those for the F480M filter and thus lie behind those points.}
 \label{fig:depths}
\end{figure}

Considering the above, the choice of optimal filter must also be cross-evaluated with the amount of time required to reach the required imaging depths. To do so, we use the JWST Exposure Time Calculator (ETC) to estimate the amount of time required to get to a 5$\sigma$ depth of $\sim$29-29.4 AB in each of the four $>4 \mu$m filters. We perform the calculations assuming either 6 or 8 groups per exposure and either 1 or 3 integrations per exposure and a ``DEEP8'' read mode. The resulting median (interpolated) depths and 1$\sigma$ standard deviations are presented in Figure \ref{fig:depths}. For each of the four $>4 \mu$m filters to reach the aforementioned depth, we find rough estimates of $\sim$10-20 ks ($\sim$3-5 hrs), $\sim$20-40 ks ($\sim$5-11 hrs) and $>$40 ks ($>$11 hrs) of required observing time for the F410M, F430M and F460M/F480M filters, respectively. These results clearly illustrate the delicate balance between opting for filters that yield the most significant gains (i.e., F430M) or those which require less exposure time (e.g., F410M); indeed, with a relatively constant F430M/F410M exposure time ratio of $\sim2.5$, the difference in overhead and required exposure time for depths less than $\sim28.5$ AB are modest, while these differences become significantly amplified for deeper imaging. Inevitably, the ultimate choice of filter will depend critically on the required depth, target and strategy of choice.

For a comparison with spectroscopic capabilities using NIRCam and NIRSpec (MIRI samples wavelengths redward of 5 $\mu$m and thus is not suitable for a comparison here), we use the star-forming galaxy spectrum shown in Figure \ref{fig:sedfilters} at $z=8.4$ (normalised to m$_{F200W}\sim$26 AB and with EW([\OIII]+H$\beta\sim$1300 \AA)) and simulate various spectroscopic observations of similar times to the NIRCam imaging. Assuming a similar setup and exposure time to those above, NIRCam grism observations with the F430M (or F444W) filter would be unable to gain any significant detection. On the other hand, for fixed slit and MOS observations, a total of 32.6 ks (i.e., $\sim$9 hrs, assuming the G395M grating and slit dimensions of 0.4$''\times''$3.8$''$) would be sufficient only to obtain $\sim1.8\sigma-2.8\sigma$ detections across the continuum blueward of 4 $\mu$m but $\sim5\sigma$ detections of weak nebular emission lines and $>5\sigma$ detections of strong lines such as [\OII], H$\beta$ and both [\OIII] lines. Fortunately, NIRSpec Prism capabilities offer far more sensitive spectroscopic observations, with only $\sim$2 hrs required to gain $>3\sigma$ continuum detections across all wavelengths blueward of 4 $\mu$m, and $\sim8\sigma-20\sigma$ detections of the [\OIII]+H$\beta$ lines. However, it is important to note that while using fixed slit or MOS observations with NIRSpec may afford high S/N spectra, they do not afford the same sky coverage as NIRCam imaging, and thus are better placed for the characterisation of known objects rather than the discovery and simultaneous characterisation of galaxy populations across large (and potentially unknown) areas of the sky.

\section{Conclusions}
We have conducted an analysis of galaxy parameter recovery and accuracy through mock \jwst/NIRCam galaxy photometry and SED-fitting, at redshifts $z\sim7-11$. While future wide-band imaging from ERS and GTO programs will undoubtedly offer unprecedented depth and quality at NIR wavelengths, their large bandwidths prevent the isolation and precise characterisation of rest-frame optical galaxy properties. We find the addition of a single, red medium-band NIRCam filter to wide-band filters adopted by ERS and GTO programs (and at comparable depth) can significantly increase the accuracy of a large number of galaxy parameters (global or spatially-resolved), in particular galaxy metallicity, stellar age and mass, and emission line flux. The difference in accuracy can span up to $\sim$0.6 dex for both bright and faint galaxy populations. This is particularly true at redshifts of $z<10$.

Defining the gain in accuracy from an additional filter as the sum of the accuracy differences across all properties and magnitudes, we find the most significant gains are found for the $z\sim7-9$ galaxy populations, although more marginal gains are still found for the $z\sim10-11$ populations. We argue the addition of a single, strategically-placed medium-band filter to existing ERS and GTO data sets offers a valuable and relatively inexpensive approach towards maximising the characterisation of the $z\sim7-11$ galaxy populations, especially in light of the short lifespan of \jwst - while the final choice of filter will inevitably depend on the science goal and target, we find on a per-galaxy basis the F430M filter affords the most significant gains in accuracy for both $z\sim8$ galaxies and overall, compared to other medium-band filters. Given sufficient angular resolution and signal-to-noise, our results apply both to integrated and spatially-resolved galaxy properties, and thus the observational strategies discussed here are valid for both.

\acknowledgments
GRB and TT acknowledge financial support from NASA through grant JWST-ERS-1342. CAM acknowledges support by NASA Headquarters through the NASA Hubble Fellowship grant HST-HF2-51413.001-A awarded by the Space Telescope Science Institute, which is operated by the Association of Universities for Research in Astronomy, Inc., for NASA, under contract NAS5-26555. We thank the anonymous referee for their valuable feedback and suggestions, which improved this paper. GRB would also like to extend his thanks and appreciation to Adam Carnall for many useful discussions on the modification and use of his \texttt{Bagpipes} SED-fitting code.


\begin{thebibliography}{}
\bibitem[Behroozi et al.(2020)]{behroozi20} Behroozi, P., Conroy, C., Wechsler, R.~H., et al.\ 2020, \mnras, 499, 5702. doi:10.1093/mnras/staa3164
\bibitem[Bradley et al.(2012)]{bradley12} Bradley, L.~D., Trenti, M., Oesch, P.~A., et al.\ 2012, \apj, 760, 108
\bibitem[Bradley et al.(2014)]{bradley14} Bradley, L.~D., Zitrin, A., Coe, D., et al.\ 2014, \apj, 792, 76
\bibitem[Calzetti et al.(2000)]{calzetti2000} Calzetti, D., Armus, L., Bohlin, R.~C., et al.\ 2000, \apj, 533, 682
\bibitem[Carnall et al.(2018)]{carnall18} Carnall, A.~C., McLure, R.~J., Dunlop, J.~S., et al.\ 2018, \mnras, 480, 4379
\bibitem[Castellano et al.(2017)]{castellano17} Castellano, M., Pentericci, L., Fontana, A., et al.\ 2017, \apj, 839, 73
\bibitem[Coe et al.(2019)]{coe19} Coe, D., Salmon, B., Brada{\v{c}}, M., et al.\ 2019, \apj, 884, 85
\bibitem[De Barros et al.(2019)]{debarros19} De Barros, S., Oesch, P.~A., Labb{\'e}, I., et al.\ 2019, \mnras, 489, 2355
\bibitem[Ellis et al.(2013)]{ellis13} Ellis, R.~S., McLure, R.~J., Dunlop, J.~S., et al.\ 2013, \apjl, 763, L7
\bibitem[Endsley et al.(2021)]{endsley20} Endsley, R., Stark, D.~P., Chevallard, J., et al.\ 2021, \mnras, 500, 5229
\bibitem[Finkelstein et al. (2013)]{finkelstein13} Finkelstein, S., Papovich, C., Dickinson, M. et al. \ 2013, \nat, 502, 524
\bibitem[Grogin et al.(2011)]{grogin11} Grogin, N.~A., Kocevski, D.~D., Faber, S.~M., et al.\ 2011, \apjs, 197, 35
\bibitem[Hashimoto et al.(2018)]{hashimoto18} Hashimoto, T., Laporte, N., Mawatari, K., et al.\ 2018, \nat, 557, 392
\bibitem[Hoag et al.(2017)]{hoag17} Hoag, A., Brada{\v{c}}, M., Trenti, M., et al.\ 2017, Nature Astronomy, 1, 0091
\bibitem[Kauffmann et al.(2020)]{kauffmann20} Kauffmann, O.~B., Le F{\`e}vre, O., Ilbert, O., et al.\ 2020, \aap, 640, A67. doi:10.1051/0004-6361/202037450
\bibitem[Kemp et al.(2019)]{kemp19} Kemp, T.~W., Dunlop, J.~S., McLure, R.~J., et al.\ 2019, \mnras, 486, 3087
\bibitem[Koekemoer et al.(2011)]{koekemoer11} Koekemoer, A.~M., Faber, S.~M., Ferguson, H.~C., et al.\ 2011, \apjs, 197, 36
\bibitem[Labb{\'e} et al.(2013)]{labbe13} Labb{\'e}, I., Oesch, P.~A., Bouwens, R.~J., et al.\ 2013, \apjl, 777, L19
\bibitem[Laporte et al.(2017a)]{laporte17a} Laporte, N., Ellis, R.~S., Boone, F., et al.\ 2017, \apjl, 837, L21
\bibitem[Lotz et al.(2017)]{lotz17} Lotz, J.~M., Koekemoer, A., Coe, D., et al.\ 2017, \apj, 837, 97
\bibitem[Mainali et al.(2018)]{mainali18} Mainali, R., Zitrin, A., Stark, D.~P., et al.\ 2018, \mnras, 479, 1180
\bibitem[Oesch et al.(2014)]{oesch14} Oesch, P.~A., Bouwens, R.~J., Illingworth, G.~D., et al.\ 2014, \apj, 786, 108
\bibitem[Oesch et al.(2015)]{oesch15} Oesch, P.~A., van Dokkum, P.~G., Illingworth, G.~D., et al.\ 2015, \apjl, 804, L30
\bibitem[Oesch et al.(2018)]{oesch18} Oesch, P.~A., Bouwens, R.~J., Illingworth, G.~D., et al.\ 2018, \apj, 855, 105
\bibitem[Oke \& Gunn(1983)]{oke83} Oke, J.~B., \& Gunn, J.~E.\ 1983, \apj, 266, 713 
\bibitem[Roberts-Borsani et al.(2016)]{rb16} Roberts-Borsani, G.~W., Bouwens, R.~J., Oesch, P.~A., et al.\ 2016, \apj, 823, 143
\bibitem[Roberts-Borsani et al.(2020)]{rb20} Roberts-Borsani, G.~W., Ellis, R.~S., \& Laporte, N.\ 2020, \mnras, 497, 3440
\bibitem[Salmon et al.(2018)]{salmon18} Salmon, B., Coe, D., Bradley, L., et al.\ 2018, \apjl, 864, L22
\bibitem[Schmidt et al.(2014)]{schmidt14} Schmidt, K.~B., Treu, T., Brammer, G.~B., et al.\ 2014, \apjl, 782, L36
\bibitem[Smit et al.(2015)]{smit15} Smit, R., Bouwens, R.~J., Franx, M., et al.\ 2015, \apj, 801, 122
\bibitem[Stark et al.(2017)]{stark17} Stark, D.~P., Ellis, R.~S., Charlot, S., et al.\ 2017, \mnras, 464, 469
\bibitem[Tang et al.(2019)]{tang19} Tang, M., Stark, D.~P., Chevallard, J., et al.\ 2019, \mnras, 489, 2572
\bibitem[Tang et al.(2020)]{tang20} Tang, M., Stark, D.~P., Chevallard, J., et al.\ 2020, \mnras
\bibitem[Treu et al.(2015)]{treu15} Treu, T., Schmidt, K.~B., Brammer, G.~B., et al.\ 2015, \apj, 812, 114
\bibitem[Williams et al.(2018)]{williams18} Williams, C.~C., Curtis-Lake, E., Hainline, K.~N., et al.\ 2018, \apjs, 236, 33
\bibitem[Zitrin et al.(2015)]{zitrin15} Zitrin, A., Labb{\'e}, I., Belli, S., et al.\ 2015, \apjl, 810, L12
\end{thebibliography}

\end{document}